\documentclass[twocolumn]{aastex701}

\usepackage{graphicx}
\usepackage{hyperref}
\usepackage{aas_macros}
\usepackage{amsmath}
\usepackage{amssymb}
\usepackage{xcolor}
\usepackage{nccmath}
\usepackage{multirow}
\usepackage{subcaption}
\usepackage{gensymb}
\usepackage{soul}

\usepackage[utf8]{inputenc}
\usepackage[T1]{fontenc}

\newcommand{\Msol}{~M_{\odot}}

\begin{document}

\title{What Are Pulsar Companions Made of? Using Gravitational Tides to Probe Their Compositions}

\author[orcid=0009-0007-7272-7355]{Liam Colombo-Murphy}
\affiliation{Department of Physics, 1156 High St., University of California Santa Cruz, Santa Cruz, CA 95064, USA}
\email[show]{lvmurphy@ucsc.edu}  

\author[orcid=0009-0008-8199-580X]{Lucas Brown} 
\affiliation{Department of Physics, 1156 High St., University of California Santa Cruz, Santa Cruz, CA 95064, USA}
\affiliation{Santa Cruz Institute for Particle Physics, 1156 High St., Santa Cruz, CA 95064, USA}
\email{lubrown@ucsc.edu}

\author[orcid=0000-0002-9159-7556]{Stefano Profumo} 
\affiliation{Department of Physics, 1156 High St., University of California Santa Cruz, Santa Cruz, CA 95064, USA}
\affiliation{Santa Cruz Institute for Particle Physics, 1156 High St., Santa Cruz, CA 95064, USA}
\email{profumo@ucsc.edu}

\author[orcid=0009-0005-7861-3290]{M. Grant Roberts} 
\affiliation{Department of Physics, 1156 High St., University of California Santa Cruz, Santa Cruz, CA 95064, USA}
\affiliation{Santa Cruz Institute for Particle Physics, 1156 High St., Santa Cruz, CA 95064, USA}
\email{migrober@ucsc.edu}

\author[orcid=0009-0003-8229-0127]{Aya Westerling}
\affiliation{Department of Physics, 1156 High St., University of California Santa Cruz, Santa Cruz, CA 95064, USA}
\email{aylweste@ucsc.edu}  

\begin{abstract}
Low eccentricity, short orbital period pulsar companions may provide a probe to study novel dense and stable exoplanet internal compositions due to the potentially significant orbital evolution they experience caused by strong gravitational tides. We model the tidal characteristics such as apsidal motion constants, orbital precession, and tidal deformability for a variety of equations of state to be compared with values recovered via pulsar timing for a sample of four systems: PSR J1719-1438b, PSR J0636+5128b, PSR J2322-2650b, and PSR J1807-2459Ab. With this method, we hope to place stringent limits on the chemical and structural composition of these objects. Through limiting the internal composition of pulsar companions, we aim to elucidate their unique history and formation.

\end{abstract}


\section{Introduction}

The discovery of low-eccentricity, short period companions orbiting pulsars represented a marked shift in the understanding of which environments could produce planet-mass companions. Unlike conventional planetary systems with companions orbiting main sequence stars, these companions inhabit strong gravitational fields and experience extreme irradiation from their rapidly rotating neutron star hosts \citep{Draghis_2018}. These pulsar companions are detected via the periodic variations they induce in pulse time of arrivals (TOA) measurements. The high precision of millisecond pulsar timing methods enables the detection of companions with masses as low as those of the largest asteroids, resulting in the discovery of a wide range of companion masses orbiting pulsars \citep{Wolszczan_1997_NS-Planets}.

Since the observation of the Hulse-Taylor system, PSR B1913+16, pulsar binaries have served as rich test beds for studying general relativity and post-Newtonian dynamics \citep{Hulse-Taylor}. These systems have similarly advanced the field of planetary science, providing the first discovery of exoplanets through pulsar timing measurements in the PSR 1257+12 ``Lich'' system \citep{Lich_System_Wolszczan-Frail}. The first observations of a millisecond pulsar in PSR B1937+21 \citep{Backer_1982} (now designated PSR J1939+2134), brought forward new questions regarding the evolutionary pathways of neutron star binary systems. In order to achieve the short pulsar rotation periods, a companion is required to be ``recycled'' via accretion of the pulsar companion, transferring angular momentum and spinning up the host pulsar to a millisecond period \citep{Bhattacharya_1991}. For companions to be observed about these millisecond pulsars, the recycled donor companion may have been originally sufficiently massive to spin-up the host but reach a detached state where Roche-lobe overflow siphoning ceases before the companion is completely accreted, possibly leading to the unique objects we discuss in this paper. Because of this, the study of pulsar companions not only provides us with a way of studying novel `planetary' compositions, they also provide more insight into the history of their host millisecond pulsars. 

As an explicit example, PSR J1719-1438b is a 1.2 Jupiter-mass companion with an orbital period of approximately 2.18 hours around its host millisecond pulsar; the corresponding Roche-lobe radius is found to be $0.43 ~R_J$, implying that the minimum mean density for the companion is 20.2 g/cm$^3$ \citep{Bailes_2011}, which is about 6 times the density of diamond. This extreme density, combined with the potential for a carbon composition, has led to PSR J1719-1438b receiving the colloquial name of ``Diamond Planet'' \citep{Susobhanan_2018}.

Due to these abnormal conditions, short orbital period pulsar companions have become a useful probe for very dense and stable internal companion compositions. The significant gravitational fields experienced by these objects may hint at compositions seen in no other exoplanetary systems. Studying the compositions of these low-eccentricity, tight orbit companions provides an opportunity to understand what kind of formation mechanism could produce such systems, as well as inform us on their population size. We provide a framework that allows us to study the interior of these objects via pulsar timing observations. Through modeling the tidal observables, primarily orbital precession and decay, of a given internal composition and comparing to pulsar timing, we provide a procedure by which the structure and chemical make-up of these companions can be probed and constrained.

This work is organized as follows. First, in Section \ref{sec:Data} we discuss four objects for which our formalism will be especially relevant, and review and update what has been observed, calculated, and speculated about these systems. Next, in Section \ref{sec:model_details} we discuss the selection of the equations of state (EOS) we use to display a variety of tidal characteristics and detail the numerical methods employed by our python pipeline, APSIDE\footnote{https://github.com/LVColombo-Murphy/APSIDE.git} (A Python Solver for Integrating tiDal characteristics from Equations of state), to model these tidal characteristics. In Section \ref{subsec:modeling_apsidal_motion}, we discuss the primary method of studying the companion's interiors, showing that the compact nature of the host pulsar can potentially lead to clean measurements of apsidal motion and, therefore, constrain the companion's apsidal motion constant $k_{2,c}$. We also touch on the secular decay of the orbit, $\dot{P}_b$, which can be difficult to isolate tidal contributions from, but may provide another method of probing the internal behavior of these objects. In Section \ref{sec:pulsar_timing}, we give a description of our use of the pulsar timing software, PINT, and how we might differentiate between compositions with pulsar timing. Following this, we describe the results of our pipeline APSIDE in Section \ref{sec:results} as well as how feasible these methods are for precise measurements of companion internal characteristics and provide a general discussion of our methods in Section \ref{sec:conclusion}. In Appendix \ref{subsec:tidal_deformability}, we discuss the feasibility of gravitational wave observations of these systems for various compositions.

%%%%%%%%%%%%%%%%%%%%%%%%%%%%%%%%%%%%%%%%%%%%%%%%%%%%%%%%%%%%%%%%%%%%%%%%%%%%%%%%%%%%
%%%%%%%%%%%%%%%%%%%%%%%%%%%%%%%%%%%%%%%%%%%%%%%%%%%%%%%%%%%%%%%%%%%%%%%%%%%%%%%%%%%%

\section{Data/Background}\label{sec:Data}

We have chosen to highlight a sample of 4 systems to which our framework can be applied, spanning masses ranging from sub-Jupiter to super-Jupiter. Due to their low eccentricity, small orbital radii, and compact pulsar hosts, these companions exhibit abnormally high minimum densities. For each system, we give a brief overview of what is known and inferred from observation, as well as provide prominent literature on these objects. We update previous Roche-lobe radii estimations with the Eggleton approximation \citep{Eggleton_1983}, 
\begin{gather}\label{Roche-lobe}
    R_L = \frac{0.49q^{2/3}a}{0.6q^{2/3}+\ln(1+q^{1/3})},
\end{gather}

\noindent where $a$ is the binary separation, $q=M_c/M_p$ is the binary mass ratio, $M_c$ is the companion mass, and $M_p$ is the pulsar mass. This expression is accurate to within 1\% for all $q$ and more precise than the well known Paczynski Approximation which is used in some prior literature \citep{Eggleton_1983, Paczynski_1971}. Unless otherwise specified, we assume a canonical mass of the neutron star hosts of $M_p = 1.4\Msol$, as is standard in the literature \citep{Bailes_2011, Stovall_2014, Spiewak_2017, Ransom_2001}.

\subsection{PSR J1719-1438b}\label{subsec:J1719}
Orbiting a 5.8 millisecond pulsar, PSR J1719-1438b has a period of 7837 seconds and minimum mass of $1.22~M_J$ with a resulting Roche-lobe of $0.43~R_J$ \citep{Bailes_2011,van_Haaften_2012}. No measurement of the orbital period derivative $\dot{P}_b$ has yet been made. Interestingly, this system is `detached', meaning the companion is no longer Roche-lobe filling and that there is currently no mass transfer between donor and host. Therefore, we know that the mean density of the companion is higher than 20.2 g/cm$^3$ because the radius is necessarily smaller than the Roche-lobe.

Simulations of Ultra-Compact X-Ray Binaries (UCXB) have shown that the characteristics of PSR J1719-1438 can be well modeled by evolving a system where the companion experiences heating from its host, bloating its radius and causing evaporating wind, both probable events that allow the system to feasibly form within a Hubble time \citep{Benvenuto_2012, van_Haaften_2012}. These simulations were performed for the rest of our sample, showing similar agreement with the observed companion masses and periods \citep{Guo_2022}.

It has also been proposed that these objects could be a nugget of cold, self-bound strange-quark matter \citep{Horvath_2012, Wang_2021}. These more exotic compositions lead to an extremely dense companion with a radius on the order of 1 km. Strange quark matter candidates would exhibit very different tidal characteristics from more conventional compositions, as discussed later in Section \ref{sec:results}. As such these systems may provide a quick means of ruling out more exotic compositions through measurements of the apsidal motion.

\subsection{PSR J0636+5128b}
Previously referred to as PSR J0636+5129b, updated measurements show that PSR J0636+5128b orbits a 2.87 millisecond pulsar with a period of 5750 seconds. First discovered in the Green Bank Celestial Cap Survey \citep{Stovall_2014}, updated observations of PSR J0636+5128b revealed that it has a mass of 19.9$M_J$, larger than the original minimum estimate of 7.4$M_J$ \citep{Draghis_2018}. We find that the Roche-lobe is $0.84 ~R_J$ with a minimum mean density of 40.1 g/cm$^3$. \citet{Kaplan_2018} inferred a mass of $17.9\pm 2.4 ~M_J$, radius of $0.76 \pm 0.14 ~R_J$, and mean density of $54 \pm 26 ~\text{g/cm}^3$, via light curve modeling, all broadly consistent with other current measurements \citep{Kaplan_2018}. We use the latest results when modeling for PSR J0636+5128b. Fitting from the NANOGrav 11 year data set provides a period derivative of $\dot{P}_b = 2.5(3)\times10^{-12} ~s~s^{-1}$ \citep{Arzoumanian_2018}.

\subsection{PSR J2322-2650b}
PSR J2322-2650b is the first pulsar companion with a confirmed atmospheric spectrum. First discovered in the High Time Resolution Universe survey, this object orbits a 3.5 ms pulsar with a period of 27904 seconds, has a minimum mass of $0.8 ~M_J$, and was found to exhibit blackbody radiation consistent with a tidally locked, near Roche-lobe filling companion \citep{Spiewak_2017}. An upper bound on the orbital period derivative was calculated to be $\dot{P}_b \leq 6\times10^{-11}~s~s^{-1}$ but no measurements have yet been made \citep{Spiewak_2017}. This object is our sample's lowest minimum mean density companion at 1.59 g/cm$^3$ with a Roche-lobe of $0.87 ~R_J$, assuming the minimum mass. Observations with James Webb Space Telescope's NIRSpec found that, PSR J2322-2650b's atmosphere was depleted of hydrogen and helium but remain carbon rich, challenging the formation theories of UCXB's producing Helium-rich objects through Roche-lobe overflow and evaporating winds \citep{Zhang_2025}. This observation has supported the idea that these objects are some kind of ``Diamond Planet''.

\subsection{PSR J1807-2459Ab}
Simultaneously discovered by \citet{Ransom_2001} and \citet{DAmico_2001}, PSR J1807-2459Ab has a period of 6142 s around its 3.06 ms pulsar host and has an inferred minimum mass of $9.4 ~M_J$. This leads to a Roche-lobe of $0.71 ~R_J$ and a minimum density of 34.4 g/cm$^3$. This system has been observed to have an orbital period derivative of $\dot{P_b} = -1.142(62) \times10^{-12}~s~s^{-1}$ \citep{Lynch_2012}. The measurement of $\dot{P}_b < 0$ implies that this system is possibly the only in our sample in which tidal contributions to the orbital period derivative dominate, potentially giving another method of probing the interior of this object.

%%%%%%%%%%%%%%%%%%%%%%%%%%%%%%%%%%%%%%%%%%%%%%%%%%%%%%%%%%%%%%%%%%%%%%%%%%%%%%%%%%%%
%%%%%%%%%%%%%%%%%%%%%%%%%%%%%%%%%%%%%%%%%%%%%%%%%%%%%%%%%%%%%%%%%%%%%%%%%%%%%%%%%%%%

\section{Model Framework}\label{sec:model_details}

In this section, we present the equations of state, stellar structure framework, and tidal formalism that underlie our modeling of pulsar companion interiors and their observable orbital effects.

\subsection{Equations of State}\label{subsec:EOS}

We provide a survey of the EOS we used to test our method to differentiate compositions via tidal characteristics. The compositions we study vary from well-known planetary EOS to more exotic EOS such as the MIT Bag Model. For carbon-oxygen compositions, we used the model provided in \citet{Podolak_2023}. We similarly use \citet{Zapolsky_1969} for cold carbon and cold carbon-oxygen (50\% carbon 50\% oxygen by mass) to probe extreme density carbon compositions as potential proxies for Jupiter mass CO stripped white dwarf remnants. We use the EOS for pure hydrogen, pure helium, and H-He with 25\% helium by mass from \citet{Zapolsky_1969, Salpeter_1967}. To show how more standard terrestrial compositions would behave in these systems, we use the well known iron, magnesium silicate, silicon carbide, and H$_2$O EOS from \citet{Seager_2007}. We use the white dwarf EOS in \citet{Sagert_2006} which becomes an $n=1.5$ polytrope at low masses in the non-relativistic limit. For the last of the companion compositions, we test the MIT Bag model from \citet{Chodos_1974}. We model our host pulsars with the SLy EOS from \citet{Douchin_2001}. 

With this array of EOS, we cover a range of interior characteristics with APSIDE and provide a visualization of our framework while remaining agnostic about the model details/implementation. With future increases in the amount of pulsar TOA measurements, we will be able to better differentiate between the various EOS. It is important to note that our current modeling does not account for more dynamic interior processes such as convection or strong day-side heating of the companion, leading to internal anisotropies and possibly deep penetrating convection \citep{Conrad_Burton_2023}. A rigorous study of these effects in our sample of pulsar companion systems is yet to be done, but for the class of larger mass pulsar companions called black widows, these effects have been studied \citep{Ginzburg_2020, Romani_2016, Voisin_2020}. We expect that these processes will alter the tidal characteristics, but leave accounting for these effects to future works.

\subsection{Determining the Apsidal Motion Constant $k_2$}\label{subsec:how_to_get_k2}

For our purposes, the most useful probe of interior composition is to obtain the apsidal motion constant $k_{2,{\rm aps}}$ of the companion \citep{Love_1909, Sterne_1939_apsidal-k2}. The apsidal motion constant is a dimensionless quantity, ranging from 0-0.75, which encodes the radial distribution of mass in an object as well as the linear response of the body's quadrupole to a gravitational tide or perturbation \citep{Love_1909}. A $k_{2,{\rm aps}}$ of 0.75 indicates a constant density throughout the interior and corresponds to the most tidally deformable body, while an object that has the most extreme radial mass concentration, a black hole, has a $k_2$ of 0 \citep{Iteanu_2025}.

Note that $k_2$ also often denotes the second-order tidal Love number, which we will denote as $k_{2,{\rm Love}}$. These two numbers simply differ by a factor of 2 as follows: $k_{2,{\rm Love}} = 2k_{2,{\rm aps}}$. In stellar literature, it is very common for $k_{2,{\rm aps}}$ to be referred to as ``the Love Number $k_2$''. This unfortunate historical overlap of notation has been noted and discussed in other papers \citep{Ragozzine_Wolf_2009, Csizmadia_Hellard_Smith_2019}. In this work, a $k_2$ without subscript of "Love" denotes the apsidal motion constant $k_{2,{\rm aps}}$.

In the following sections we will discuss the derivations of the apsidal motion constant $k_2$ used by our pipeline APSIDE in two regimes: the non-relativistic regime through the formalism developed in \citet{Sterne_1939_apsidal-k2}, and the relativistic regime following the methods developed in \citet{Hinderer_2008, Postnikov_2010}.

\subsubsection{Hydrostatic Equilibrium Equations}\label{sec:Hydrostatic_TidalDeform} 

In order to calculate $k_{2}$ for a planet, we use the Hydrostatic Equilibrium equations,

\begin{eqnarray}
    &&\frac{dP}{dr} = -\frac{Gm(r)\rho (r)}{r^2},\\
    &&\frac{dm}{dr} = 4 \pi  r^2 \rho(r),
\end{eqnarray}

\noindent for which an equation of state for the object of study must be specified,

\begin{align}
    P(r) = f(\rho(r), T(r)),
\end{align}

\noindent where $f$ is a function which relates various thermodynamic variables which are unique to a given material or species of object. 

By integrating the Hydrostatic Equilibrium equations from the center of a planet to the surface, using the central boundary conditions $m(0)=0$, $P(0)=P_c$, and $\rho(0)=\rho_c$, we are able to solve for the radial dependencies for pressure and mass. The radial profile of the density can then be found using an EOS's density-pressure relation $\rho(r)=\rho(P(r))$. To find $k_{2}$, we use the formalism developed in \citet{Sterne_1939_apsidal-k2},

\begin{equation}
    k_{2} = \frac{1}{2} \left( \frac{3-\eta_2(R)}{2+\eta_2(R)} \right).
\end{equation}

\noindent Where $\eta(R)$ is the value of the logarithmic derivative of the metric perturbation at the surface radius $R$, which is the solution of the following differential equation:

\begin{align}\label{eq:etadifferential}
    r \frac{d\eta_2}{dr} + \eta_2^2 - \eta_2 - 6 + \frac{6\rho(r)}{\rho_m(r)}(\eta_2 + 1) = 0~,
\end{align}

\noindent where $\rho_m(r) = \frac{m(r)}{4/3 \pi r^3}$, the average density within a shell of radius $r$. The boundary condition for this equation is $\eta_2(0) = 0$. When calculating $k_{2,{\rm Love}}$ the factor of $1/2$ can be removed from Equation \eqref{eq:etadifferential} as seen in \citet{Becker_2013}. From $k_{2}$, we find the tidal deformability $\lambda$ of the body,

\begin{equation}\label{eq:tidallambda}
    \lambda =  \frac{2R^5}{3G}k_{2}~,
\end{equation}

\noindent as well as the dimensionless tidal deformability with $M = m(R)$ \citep{Hinderer_2010},

\begin{equation}\label{eq:dimensionless_tidallambda}
    \Lambda = k_{2} \left(\frac{Rc^2}{GM} \right)^5 = \frac{\lambda c^{10}}{G^4 M^5}~.
\end{equation}

\noindent A discussion of the tidal deformability and dimensionless tidal deformability is provided in Appendix \ref{subsec:tidal_deformability}.

\subsubsection{TOV Equations}\label{sec:TOV_TidalDeform}

It is necessary to switch from the Hydrostatic Equilibrium equations to the Tolman-Oppenheimer-Volkoff (TOV) equations when studying relativistic internal compositions \citep{Oppenheimer-Volkoff_1939,Tolman_1939}. In this regime it becomes much easier to use geometrized units where $G=c=1$,

\begin{eqnarray}\label{TOVequations}
    &&\frac{dP}{dr}=-\frac{m(r) \epsilon(r)}{r^2} \left[1+\frac{P(r)}{\epsilon(r)} \right] \left[1+\frac{4\pi P(r) r^3}{m(r)} \right]  \nonumber \\ && \quad \qquad \times \left[1-\frac{2m(r)}{r}\right]^{-1} , \\
    % &&=-\left( \epsilon(r)+P(r) \right) \frac{m(r) + 4\pi P(r) r^3}{r \left( r-2m(r) \right)} \\
    &&\frac{dm}{dr}=4\pi r^2 \epsilon(r).
\end{eqnarray}

Here $\epsilon(r)=\rho(r)c^2$ is the energy density profile, $P(r)$ is the radial pressure profile, and $m(r)$ is the mass enclosed within a sphere of radius r, centered at the star's core. Like in the Hydrostatic Equilibrium equations, we integrate from the core to the surface with $m(0)=0$, $P(0)=P_c$, and $\epsilon(0)=\epsilon_c$. As before, this provides the radial profiles for pressure and mass enclosed within a spherical shell, and energy density can be found with $\epsilon(r)=\epsilon(P(r))$. Next, we use the formalism developed in \citet{Hinderer_2008} and \citet{Postnikov_2010} to solve for the the relativistic analog of $\eta_2(R)$, $y=y(R)$, which is the solution to the following differential equation at the surface radius $R$:

\begin{eqnarray}\label{eq:ydifferential}
    &&ry'(r) y(r)^2 + y(r)e^{\lambda(r)} \left[ 1+4 \pi r^2(\rho(r) + P(r)) \right] \nonumber \\ && \qquad \qquad+ r^2Q(r) = 0, \\
    &&e^{\lambda(r)} = \left[1- \frac{2m(r)}{r} \right]^{-1},  \\
    &&\nu'(r) = 2 e^{\lambda(r)} \frac{m(r) + 4\pi P(r) r^3}{r^2},\\
    &&Q(r) = 4\pi e^{\lambda(r)} \left( 5\rho(r) + 9P(r) + \frac{\rho(r) + P(r)}{c_{s}^2(r)} \right) \nonumber \\ && \qquad \qquad- 6\frac{e^{\lambda(r)}}{r^2} - (\nu'(r))^2.
\end{eqnarray}

Here $c_s(r) = \sqrt{\frac{dP(r)}{d\epsilon(r)}}$ is the sound speed within the body at radius r and  Equation \eqref{eq:ydifferential} has the boundary condition that at the center of the body $y(0)=2$ \citep{Hinderer_2010, Wang_2021}. Note that for bare strange quark stars (SQS) described by the MIT Bag model, a small correction is necessary where $y = y(R) - 4\pi R^3 \epsilon(R)/M$. The subtracted term corrects for the non-zero energy density at the surface of the SQS \citep{Lourenco_2021,Albino_2021,Wadhwa_2025}. Using $y$, as well as the dimensionless compactness parameter $C = M/R$, where $M=m(R)$, $k_{2}$ can be calculated with from,

\begin{align}\label{eq:k2}
    k_2 = &\frac{8C^5}{5}(1 - 2C)^{2} \Big[ 2 + 2C (y - 1) - y\Big] \nonumber \\ \times &\Bigg\{2C  \Big[6 - 3y + 3C (5y - 8)\Big] \\ &+ 4C^3 \Big[13 - 11y + C (3y - 2) + 2C^2(1 + y)\Big] \nonumber \\ &+ 3(1 - 2C)^2 \Big[2 - y + 2 C(y - 1)\Big]\ln{(1 - 2 C)} \Bigg\} ^{-1}.
\end{align}

As discussed in \citep{Postnikov_2010, Wang_2021}, as the compactness becomes small ($C<0.1$), it is computationally useful to expand Equation \eqref{eq:k2} in a Taylor series,

\begin{align}\label{eq:Taylork2}
    k_{2} =& \frac{(1-2C)^2}{2} \Bigg[ \frac{(2-y)}{(3+y)} + \frac{(-6 + 6y + y^2)C}{(3+y)^2} \nonumber \\ &+ \frac{(12 - 8y + 34y^2 + y^3) C^2} {7(3+y)^3} \nonumber \\ &+ \frac{(36 + 48y + 84y^2 + 62y^3 + y^4) C^3}{7(3+y)^4} \nonumber \\
            &+ \frac{5(648 + 1476y + 1884y^2) C^4}{147(3+y)^5}
    \nonumber \\
            &+ \frac{5(1472y^3 + 490y^4 + 5y^5) C^4}{147(3+y)^5} \Bigg].
\end{align}

\noindent In the Newtonian limit as $C\rightarrow0$ \citep{Hinderer_2008, Postnikov_2010},

\begin{eqnarray}\label{eq:Newtoniank2}
    &&ry'(r) + y(r)^2 + y(r) - 6 + 4\pi r^2 \frac{\epsilon(r)}{c_s(r)^2} = 0 ,  \\
    &&k_{2} = \frac{1}{2} \left( \frac{2-y}{y+3} \right),
\end{eqnarray}

\noindent where the tidal deformability can be calculated the same as before, by using Equation \eqref{eq:tidallambda} and Eq.~\eqref{eq:dimensionless_tidallambda}. 

\subsection{ Apsidal Motion}\label{subsec:modeling_apsidal_motion}

The primary means we have of indirectly observing an object's $k_2$ is through pulsar timing measurements of the apsidal motion, the precession of the companion's orbit. The precession of the apsides, the furthest and closest points of the orbit around the host, arises due to 3 main contributions: the General Relativistic precession and the two components of Newtonian precession \citep[from classical spin-orbit coupling and static tidal interactions, see ][]{Susobhanan_2018}. The GR contribution is modeled as follows:

\begin{align}\label{eq:wdot_GR}
    \dot{\omega}_{\rm GR} = \frac{3G^{2/3}~n^{5/3}~(M_p+M_c)^{2/3}}{c^2 \left(1-e^2 \right)}~,
\end{align}

\noindent where $n=2 \pi /P_b$ is the mean motion, $P_b$ is the period of the binary orbit, $M_p$ is the pulsar mass which we set to the canonical $1.4\Msol$, and $M_c$ is the companion mass \citep{Barker_1975a}. We model the Newtonian rate of secular motion with the following formulae developed in \citet{Willems_2008,Bulut_2017},

\begin{align}
    \frac{\dot{\omega}_N}{360} = \xi_p k_{2,p} + \xi_c k_{2,c},
\end{align}

\noindent where the coefficients $\xi_{p}$ and $\xi_{c}$ can be calculated via the following equation with,

\begin{eqnarray}\label{eq:apsidal_coeffs}
    &&\xi_c = n ~\left( \frac{R_c}{a}  \right)^5 \Bigg[ \left( \frac{\Omega_{r,c}}{n} \right)^2 \left( 1 + \frac{M_{p}}{M_c} \right)f(e) \nonumber \\ 
    && \qquad+ 15\frac{M_{p}}{M_c} g(e)\Bigg], \\
    &&f(e) = \frac{1}{(1-e^2)^2}, \nonumber \\
    &&g(e) = \frac{(8+ 12e^2 + e^4) f(e)^{5/2}}{8}. \nonumber
\end{eqnarray}

\noindent $R_c$ denotes the object radius, $\Omega_{r,c}$ the rotational angular velocity, $a$ the binary separation, and $e$ the orbital eccentricity. In order to find $\xi_p$, take Equation \eqref{eq:apsidal_coeffs} and swap all subscripts of $p$ and $c$, for pulsar and companion respectively. With this we can find the observed mean value of $k_2$ for the two objects in the binary,

\begin{align}\label{eq:k2_obs}
    \overline{k}_{2,obs} = \frac{1}{\xi_c + \xi_p}\frac{\dot{\omega}_N}{360}
\end{align}

\noindent Ordinarily, this method is ill-suited for binaries with mass ratios $q = M_c/M_p\not\approx 1$ \citep{Dimoff_2023}; however, in the case of low eccentricity, short orbital period pulsar companions, the tidal effects of the pulsar are negligible and dominated by those of the companion \citep{Susobhanan_2018}. For modeling the host pulsar, we used the SLy EOS from \citet{Douchin_2001}, providing us the canonical neutron star of mass $1.4\Msol$ at a radius of $\approx 11.7$ km and $k_2\approx0.07$, consistent with \citet{Hinderer_2010}. From this, it is clear that the coefficient $\xi_p$ is exponentially suppressed by the $(R_p/a)^5$ term in the front. Due to the small $k_2$ of the pulsar and the extreme suppression from the $\xi_p$ term, the pulsar's contributions to the Newtonian apsidal precession are negligible. Thus, we can succinctly write the Newtonian precession in terms of its contributions from quadrupole tidal interactions and spin-orbit coupling,

\begin{eqnarray}
    &&\dot{\omega}_{\text{tidal}} = 15n~k_{2,c} \left( \frac{R_c}{a} \right)^5 \left( \frac{M_p}{M_c} \right)g(e), \\
    &&\dot{\omega}_{\text{spin}} = n~ k_{2,c} \left( \frac{R_c}{a} \right)^5 \left( 1 + \frac{M_p}{M_c} \right) \left( \frac{P_b}{P_c} \right)^2 f(e),
\end{eqnarray}

\noindent where $P_c$ is the rotational period of the companion. In our calculations, we assume that the companion is tidally locked, hence, $P_c$ is equal to the orbital period $P_b$. We make this assumption because measuring the companion's rotational period is extensively challenging; however, assuming tidal locking is a reasonable and physically motivated first approximation made throughout the relevant literature \citep{Spiewak_2017,Susobhanan_2018}. With a measurement of apsidal motion, we can now redefine Equation \eqref{eq:k2_obs} without the negligible contributions from the host pulsar, essentially allowing for a direct measurement of the apsidal motion constant of the companion,

\begin{gather}\label{eq:k2_c_obs}
    k_{2,c} = \frac{\dot{\omega}_{\text{obs}} - \dot{\omega}_{GR}}{360\xi_c}.
\end{gather}
 
\noindent Here, the observed apsidal motion is the sum of all smaller contributions, $\dot{\omega}_{\text{obs}} = \dot{\omega}_{GR} + \dot{\omega}_{\text{tidal}} + \dot{\omega}_{\text{spin}}$. This provides the primary method of comparing our modeled companions to the data from pulsar timing measurements, allowing us to study and constrain various compositions via their $k_2$ and tidal characteristics.

\subsection{Binary Period Derivative}

Another possible means of observing the companion's tidal characteristics is through the secular decay of the binary orbital period, $\dot{P}_b$, also called the binary period derivative. In most compact binary systems, like neutron star binaries, the dominant source of the secular decay is the emission of gravitational waves. However, in the systems we study, the mass-ratio of a planetary mass companion orbiting a solar mass host neutron star leads to a suppressed contribution to the binary period derivative from GW emission and may allow for a dominant contribution from tidal effects. We model the secular decay caused by the emission of GW following the methods of \citet{Damour_1983} and \citet{Blanchet_2001},

\begin{eqnarray}\label{eq:GW_binary_period_derivative}
    \dot{P}_{b,GW} = -\frac{192 \pi}{5c^5} \left( \frac{2\pi G}{P_b} \right)^{5/3} \frac{M_p M_c}{(M_p + M_c)^{1/3}} \nonumber \\ \times \frac{1 + \frac{73}{24}e^2 + \frac{37}{96}e^4}{(1-e^2)^{7/2}}.
\end{eqnarray}

\noindent The tidal contribution to the binary period derivative for a synchronously rotating or tidally locked companion is given by \citep{Peale_1978, Peale_1986},

\begin{eqnarray}
    &&\frac{dE}{dt} = -21 M_p n^3 a^2 e^2 \frac{k_{2,c}}{Q} \left( \frac{R_c}{a} \right)^5, \\
    &&E_{\rm orb} = -\frac{G M_p M_c}{2a} \Rightarrow \frac{dE}{dt} = \frac{G M_p M_c}{2a^2} \frac{da}{dt}, \\
    &&\frac{\dot{P}_b}{P_b} = \frac{3}{2}\frac{\dot{a}}{a}, \\ 
    &&\dot{P}_{b,{\rm tide}} = -126\pi e^2 \frac{M_p}{M_c}\frac{k_{2,c}}{Q} \left( \frac{R_c}{a} \right)^5,
\end{eqnarray}

\noindent where $Q$ is the companion's tidal quality factor, which can take a variety of values depending on the composition, ranging from $10-10^9$ for terrestrial planets to degenerate white dwarfs, respectively \citep{Burkart_Quataert_2013}. The tidal quality factor is a dimensionless parameter encoding how efficiently a body dissipates tidal energy; a small $Q$ denotes a strong dissipation and large $Q$ denotes a weak dissipation \citep{Goldreich_1966}. $Q$ is defined as,

\begin{align}
    Q = \frac{2\pi E_0}{\oint -\frac{dE}{dt}dt},
\end{align}

\noindent where $E_0$ is the maximum tidal energy stored in the time period over which the contour integral is taken. Depending on the value of $Q$, the relative weight of the tidal contributions to the secular decay of the orbit can be several orders of magnitude larger (smaller) than the GW contribution when $Q$ is small (large). Introducing a new unknown variable, the tidal quality factor $Q$, means this method of probing the tidal characteristics is slightly more difficult than just measuring apsidal motion. Increasing $Q$ or decreasing $k_{2,c}$ have degenerate effects on the secular decay caused by tides, $\dot{P}_{b,{\rm tide}}$, thus it does not exactly eliminate uncertainties in variables such as the companion radius or apsidal motion constant. 

Another of the difficulties with measuring the binary period derivative $\dot{P}_b$ is the degree of pollution it faces from contributions other than gravitational wave radiation and tidal effects. For a detailed account of the many phenomena that can alter the measurement of the binary period derivative, see \citep{Shklovskii_1970, Damour_1991}. These papers provide useful methods for accounting for various effects that might lead to incorrect measurement of $\dot{P}_b$. Due to these challenges, our primary method of studying these companions remains a measurement of apsidal motion because it provides a cleaner means of obtaining $k_{2,c}$. However, a clean measurement of the binary period derivative would provide another axis of studying overall tidal effects.

%%%%%%%%%%%%%%%%%%%%%%%%%%%%%%%%%%%%%%%%%%%%%%%%%%%%%%%%%%%%%%%%%%%%%%%%%%%%%%%%%%%%
%%%%%%%%%%%%%%%%%%%%%%%%%%%%%%%%%%%%%%%%%%%%%%%%%%%%%%%%%%%%%%%%%%%%%%%%%%%%%%%%%%%%

\section{Constraining Companion Composition using Pulsar Timing}\label{sec:pulsar_timing}

As noted in the previous sections, the orbital motion of a companion object is sensitive to its internal composition through Newtonian apsidal precession effects. This raises the distinct possibility of using the observed motion of a companion object to determine its identity, or at least pointing towards a particular class of object. Luckily, the motion of these companions can indeed be determined to remarkable accuracy via pulsar timing methods.

In \citet{Susobhanan_2018}, the authors explored this very possibility, noting that some ``diamond planet'' companion scenarios can produce Newtonian contributions to the apsidal precession rate which dwarf the purely relativistic terms. Without any Newtonian contributions, they find that the apsidal precession should be confidently detected within an observing span of less than 50 years for at least six known binary pulsar systems. With the inclusion of Newtonian contributions, the number of systems which meet this criteria could increase dramatically, and the necessary observation timescales could be significantly reduced.

In this study, we expand upon the periastron advancement modeling problem explored in \citet{Susobhanan_2018}. In particular, we aimed to examine the timescale in which one should expect to obtain a measurement of $\dot\omega$ precise enough to rule out different classes of companions. Answering this question relies on utilizing the relationships given in Equations \eqref{eq:wdot_GR}-\eqref{eq:k2_c_obs} connecting a pulsar-timing measurement of $\dot\omega_{obs}$ and the binary's other various properties to an estimate of $k_{2,c}$. We used the pulsar-timing package PINT \citep{Luo:2020ksx, Susobhanan:2024gzf} to generate mock pulsar time-of-arrival (TOA) data for the binary systems J1719-1438 and J0636+5128 spanning from 1 to 200 years in total observing baseline, with various values for $\dot\omega$ injected into the timing profile for each binary. By using PINT to subsequently analyze the simulated TOAs and recover timing solutions for each system, we were able to assess the precision with which we might expect pulsar timing methods to determine $\dot\omega$, and therefore $k_{2,c}$, as a function of time. 

Our PINT simulations utilized a 45 day observation cadence, with 50 TOA measurements occurring within a 60 minute window during each observation. The exact timing of each TOA measurement is slightly randomized in order to introduce a more realistic variation in the exact measurement cadence. Common white noise with a period of 0.1$\mu s$ is also introduced into the TOAs, consistent with order-of-magnitude estimates for the white noise present in modern pulsar timing experiments \citep{EPTA:2015ike}. Our pulsar binaries were modeled and recovered assuming an ELL1k timing model which incorporates the effect of apsidal precession \citep{Susobhanan_2018}.

\begin{figure}
    \centering
    \includegraphics[width=0.8\linewidth]{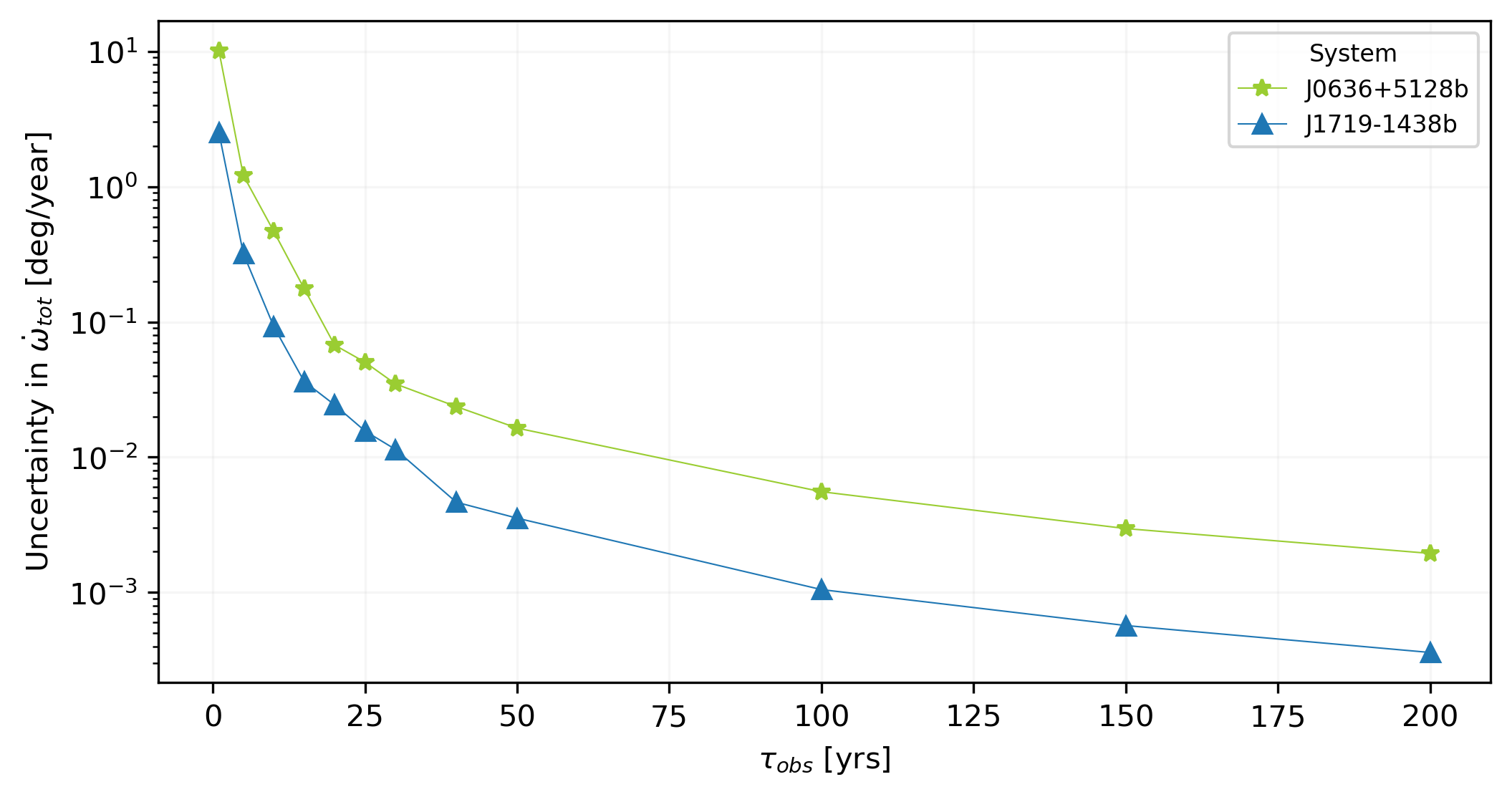}
    \caption{Uncertainties on $\dot\omega$ values recovered by PINT for our two pulsar-timing systems of special interest, J1719-1438b and J0636+5128n, as a function of years of consistent pulsar-timing measurements.}
    \label{fig:pulsar_binary_uncertainties}
\end{figure}

In Figure \ref{fig:pulsar_binary_uncertainties}, we display the uncertainty on our recovered $\dot\omega$ values as a function of time for J1719 and J0636. While J1719-1438 is shown to achieve precision quicker than J0636+5128, both pulsar systems have $\dot\omega$ values that can be determined to $\mathcal{O}(10^{-2})$-level precision within a few decades of consistent observation.

When working backwards from observations of $\dot\omega$ to a determination of $k_{2,c}$, and therefore a constraint on the composition of pulsar companions, one ideally needs to account for uncertainties associated with other parameters in equations \eqref{eq:wdot_GR}-\eqref{eq:k2_c_obs}. In most cases, the dominant source of uncertainty in these equations comes from the $(R_{c}/a)^{5}$ terms in $\dot\omega_{\rm tidal}$ and $\dot\omega_{\rm spin}$. As noted in \citet{Susobhanan_2018}, even a 50\% change in our estimate of $R_{c}$ can lead to a factor of 32 change in the Newtonian apsidal precession. Because it will likely remain difficult to precisely constrain $R_{c}$ via observation, even incredibly precise measurements of the other parameters in these equations are insufficient to determine $k_{2,c}$ to any useful level without the assumption of some particular mass-radius relationship. Further, many measurements of these systems are subject to large degeneracies between the companion orbit's inclination angle and its mass. For this reason, we choose an approach here which takes the minimum mass in each system to be its fiducial mass, noting that tighter constraints on these parameters will of course be necessary to make the methods outlined in this paper viable in the future. With a companion mass chosen, we can use various proposed EOS to derive explicit predictions for the companion's apsidal motion constant $k_{2,c}$, which we can further translate to estimations of their total apsidal precession. With these predictions for $\dot\omega_{\rm tot}$ in hand, we aim to show which classes of companion objects may be ruled out or distinguished between once measurements of $\dot\omega_{tot}$ are sufficiently precise.

Our quantification of distinguishability is as follows: for any two EOS, we can generate distinct predictions for the apsidal motion of a companion object in a given system. Let these two predictions be called $\dot\omega_{i}$ and $\dot\omega_{j}$ respectively. We conclude that these predictions are distinguishable if the uncertainty on our recovered $\dot\omega_{\rm tot}$ is significantly less than the difference between $\dot\omega_{\rm i}$ and $\dot\omega_{\rm j}$. That is, we can compute a quantity:

\begin{equation}
S_{ij} = \frac{\dot\omega_{\rm tot,i}-\dot\omega_{\rm tot,j}}{\sigma_{\dot\omega}},
\label{eq:Pbdot_tide}
\end{equation}
for each pair of companion models we may wish to discriminate between. Equipped with distinct EOS and a sufficiently clear characterization of the pulsar companion systems of interest, we can use such a metric to systematically determine which classes of companions may or may not be detected through their apsidal motion, and which may be distinguished between. The $S_{ij}$ values for the EOS considered in this study are shown in Figure \ref{fig:sij}.

%%%%%%%%%%%%%%%%%%%%%%%%%%%%%%%%%%%%%%%%%%%%%%%%%%%%%%%%%%%%%%%%%%%%%%%%%%%%%%%%%%%%
%%%%%%%%%%%%%%%%%%%%%%%%%%%%%%%%%%%%%%%%%%%%%%%%%%%%%%%%%%%%%%%%%%%%%%%%%%%%%%%%%%%%

\section{Results}\label{sec:results}
In this section, we summarize the predictions of our modeling framework and examine how effectively different companion equations of state may be differentiated and constrained in the systems we study.

\subsection{Modeling EOS with APSIDE}\label{sec:EOS_results}
\begin{figure}
    \centering
    \includegraphics[width=0.95\linewidth]{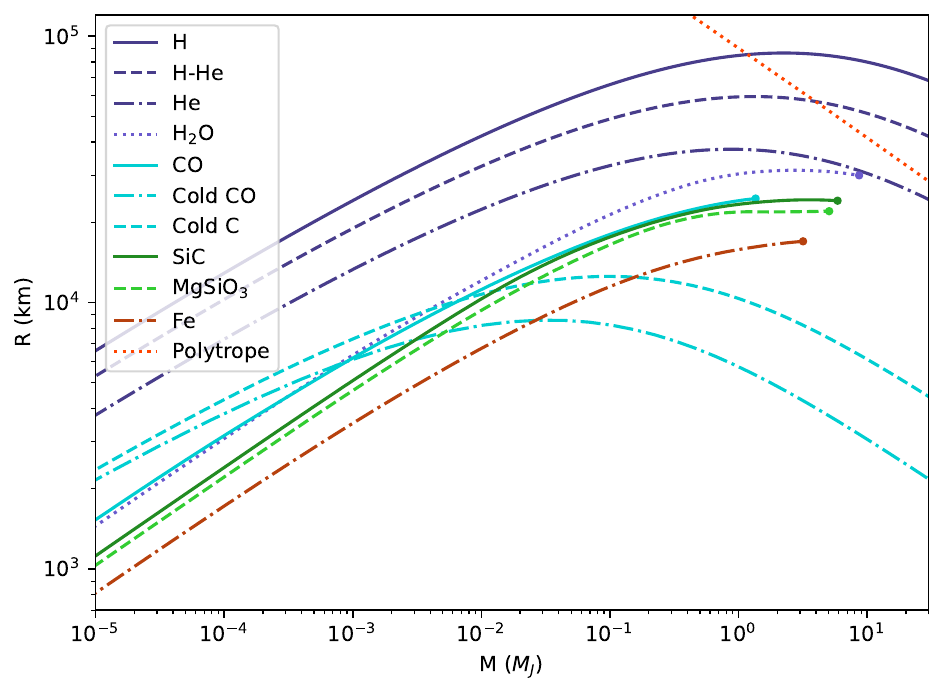}
    \caption{Plot of mass-radius relationships for companions composed of the EOS discussed in Section \ref{subsec:EOS}. Note that the curves for H$_2$O, SiC, MgSiO$_3$, and Fe end between $\sim$ 1 and 10 $M_J$, as the EOS described in \citep{Seager_2007} are only valid up for central pressures of $P_c \leq 10^{16}$ Pa. The EOS for CO is valid up to $P_c \sim 3 \cdot 10^{14}$ Pa, as described in \citep{Podolak_2023}. The label polytrope refers to an $n=1.5$ polytrope used to approximate non-relativistic white dwarfs.}
    \label{fig:Hydro_Mass-Radius_plot}
\end{figure}

\begin{figure}
    \centering
    \includegraphics[width=0.95\linewidth]{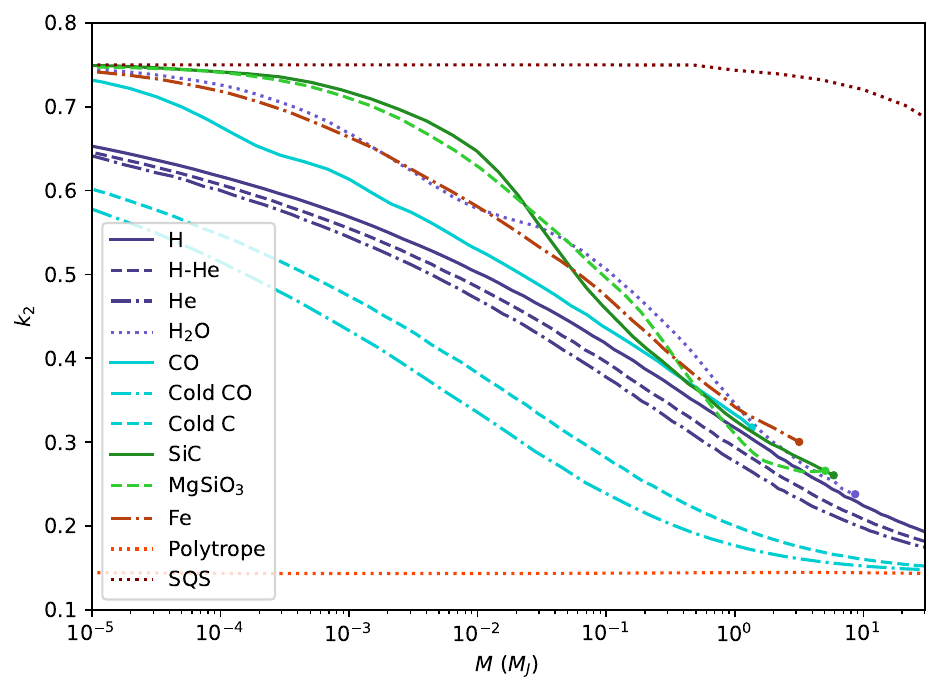}
    \caption{Mass-$k_{2}$ relationships for same EOS described in Figure \ref{fig:Hydro_Mass-Radius_plot}. Note that this is the apsidal motion constant, not the second order tidal Love number, and thus a homogeneous body has $k_{2}= 0.75$. It is interesting, but not surprising, to see that at low central pressures/masses most of these EOS tend toward a homogeneous density.}
    \label{fig:Hydro_Mass-k2_plot}
\end{figure}

\begin{figure}
    \centering
    \includegraphics[width=0.95\linewidth]{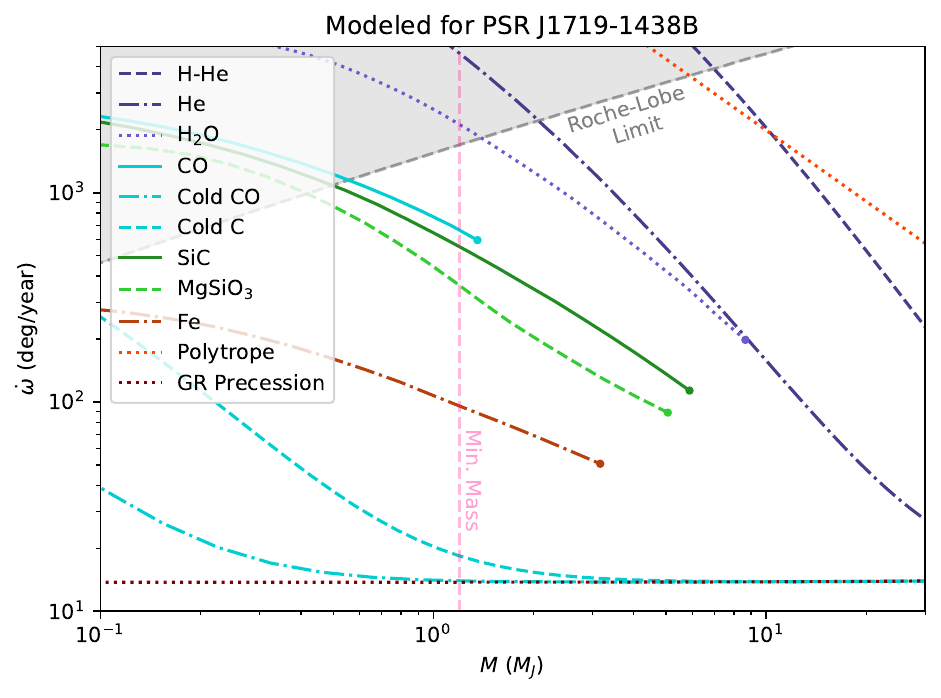}
    \caption{Plot of the relationship between mass and apsidal motion for the various internal compositions described in Section \ref{subsec:EOS}, specifically modeled to PSR J1719-1438b. The greyed out region denotes parameter-space that is ruled out by the Roche-lobe limit of this companion. The vertical pink dashed line denotes the minimum mass of PSR J1719-1438b, as described in Section \ref{subsec:J1719}. The apsidal motion returned by even the most deformable bare strange quark stars returns the GR prediction of the apsidal motion, shown in a dash-dotted sea green.}
    \label{fig:Mass-wdot_plot}
\end{figure}

\begin{figure}
    \centering
    \includegraphics[width=0.95\linewidth]{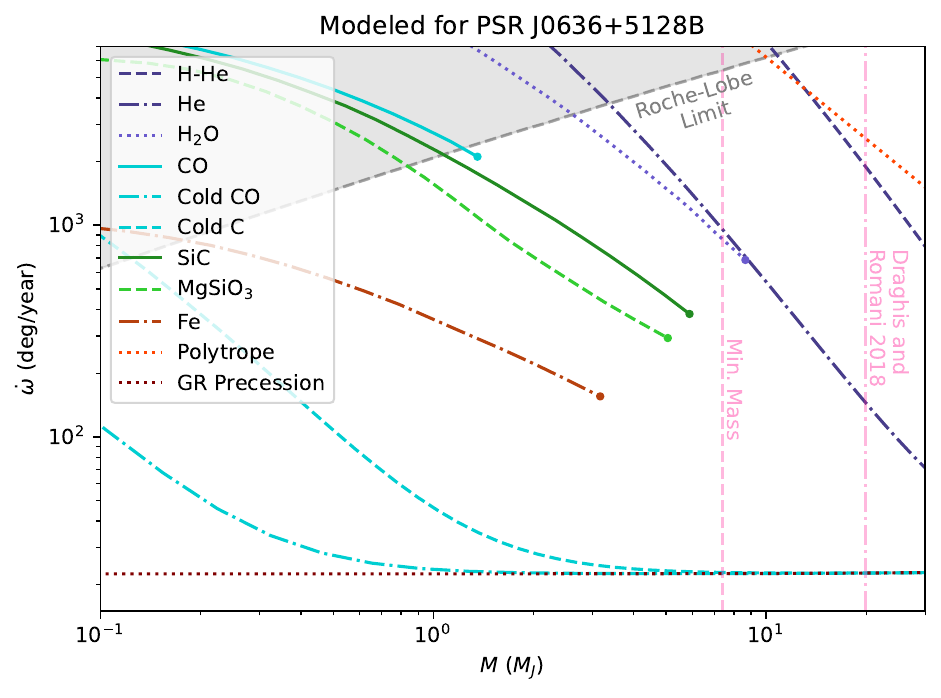}
    \caption{The same plot as Figure \ref{fig:Mass-wdot_plot} but modeled for J0636+5128b. The precession in this system can be as much as almost an order of magnitude higher than possible in J1719-1438B under the Roche-lobe limit. The additional vertical dash-dotted line denotes the measured mass of 19.9 M$_J$ from \citep{Draghis_2018}, higher than the minimum mass of 7.4 M$_J$.}
    \label{fig:Mass-wdot_plot_J0636}
\end{figure}

In this section, we discuss the results of running various EOS through our pipeline, APSIDE. First looking to the mass-radius curves in Figure \ref{fig:Hydro_Mass-Radius_plot}, with the He, H-He, H, H$_2$O, SiC, MgSiO$_3$, and Fe EOS, APSIDE precisely returns the curves from \citet{Seager_2007}. As would be expected, more gaseous compositions return much larger radii compared to the highest density compositions which yield the smallest radii. We have similarly found that APSIDE agrees with the mass radius curves for the SLy EOS seen in \citep{Hinderer_2010} and the MIT Bag Model Mass radius curves produced in \citet{Wang_2021}.

In Figure \ref{fig:Hydro_Mass-k2_plot}, we show how different compositions structure themselves internally at different masses. Recall that a body with a dense and high mass core will have a lower $k_2$ and a homogeneous density body will have a $k_2 = 0.75$. At lower masses, $k_2$ curves in this plot tend towards higher values, reaching the upper bound for some of the terrestrial compositions. At low pressures, these EOS have near constant densities as shown by Figure 3 in \citet{Seager_2007}, therefore it should be expected that the apsidal motion constant would be maximized in objects composed of these EOS with lower central pressures, and thus lower total masses. Strange quark matter objects are nearly constant density in this lower mass regime, thus remaining at a maximal $k_2$ for most of the mass range we consider. With a $n=1.5$ polytrope, we see that $k_{2} \approx 0.143$ at every mass \citep{Brooker_1955}. Polytropic equations of state assume a constant amount of radial density concentration at every pressure based on the chosen polytropic index. Thus $k_2$, which is a measure of the radial distribution of density, will be constant for any polytropic EOS. An approximation of $k_{2}$ as a function of the polytropic index was made in a previous work \citet{Hinderer_2008}, and a table of polytropic EOS apsidal motion constants is provided in \citet{Brooker_1955}. Comparing to the tabulated value, APSIDE's numerical calculation of the $n=1.5$ polytrope's $k_2$ returns less than 0.1\% error. Because apsidal motion constants have not been calculated for many non-relativistic equations of state, this check allow us to be confident in APSIDE's precision, along with the reproduction of the mass-$k_2$ curves from \citep{Hinderer_2010} and \citep{Wang_2021} for the SLy neutron star and MIT Bag Model EOS, respectively.

In Figure \ref{fig:Mass-wdot_plot}, we show how PSR J1719-1438b would precess when subject to our range of compositions. We have removed the hydrogen curve as it was almost entirely ruled out by the Roche-lobe constraint region that is shaded in grey. At PSR J1719-1438b's minimum mass of $1.2~M_J$ (shown by the vertical dashed-pink line), we see that the hydrogen-helium, helium, H$_2$O, and non-relativistic polytropic compositions are constrained by the Roche-lobe exceeding region, yet at higher masses they become viable as they no longer exceed the limit. The GR precession curve is the value of apsidal motion returned by a non-deformable, or negligibly deformable body given by Equation \eqref{eq:wdot_GR}. The highly dense Cold CO and Cold C EOS asymptotically approach this value due to their small radii and $k_2$ at higher masses. Unsurprisingly, due to their extreme densities and small radii, even the most deformable bare strange quark stars with a bag constant of $B=57$ return only the GR prediction of apsidal motion. This is true in Figure \ref{fig:Mass-wdot_plot_J0636} as well, where we model the same EOS for the system of PSR J0636+5128B. As such, the measurement of apsidal motion provides a clean method of falsifying whether these objects have more exotic compositions like strange objects. Any precession greater than the GR prediction would immediately rule out the MIT Bag Model, and other highly compact EOS.

With Figure \ref{fig:Mass-wdot_plot_J0636}, we begin to see the limitations of more solid exoplanet EOS, like those created in \citep{Seager_2007}. Due to their upper limits at central pressures of $P_c = 10^{16}$ Pa, these EOS do not reach the measured minimum mass of the companion at 7.4 M$_J$. It is our hope that as more advancements are made in EOS creation, higher pressure EOS will allow APSIDE to model solid exoplanets at higher masses.

\subsection{Feasibility of Constraining Equations of State with Pulsar Timing}\label{sec:timing_results}
As can be seen in Figures \ref{fig:Hydro_Mass-Radius_plot} and \ref{fig:Hydro_Mass-k2_plot}, many companion models which are composed of traditional baryonic matter require large radii and $k_{2,c}$, leading to the large predicted $\dot\omega_{tot}$ values seen in Figure \ref{fig:Mass-wdot_plot}. This means that even with very minimal pulsar timing data ($\mathcal{O}(1)\rm\ yr$) it should be possible to not only discriminate between particular EOS models, but to outright falsify many of them.

\begin{figure*}[t!]
\centering

\textbf{J1719$-$1438b}

\vspace{0.3em}

\includegraphics[width=0.32\textwidth]{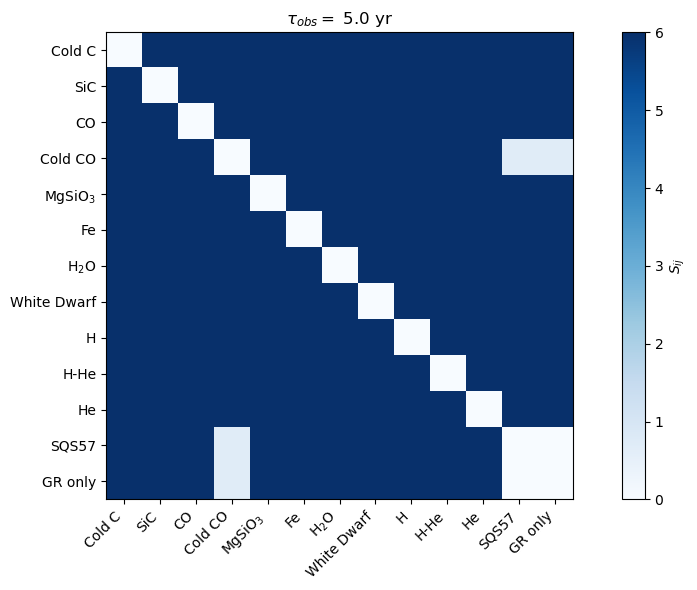}
\hfill
\includegraphics[width=0.32\textwidth]{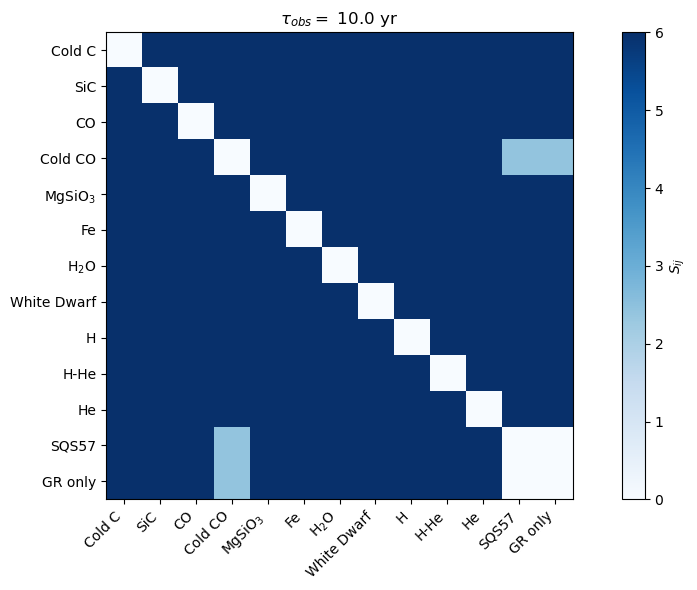}
\hfill
\includegraphics[width=0.32\textwidth]{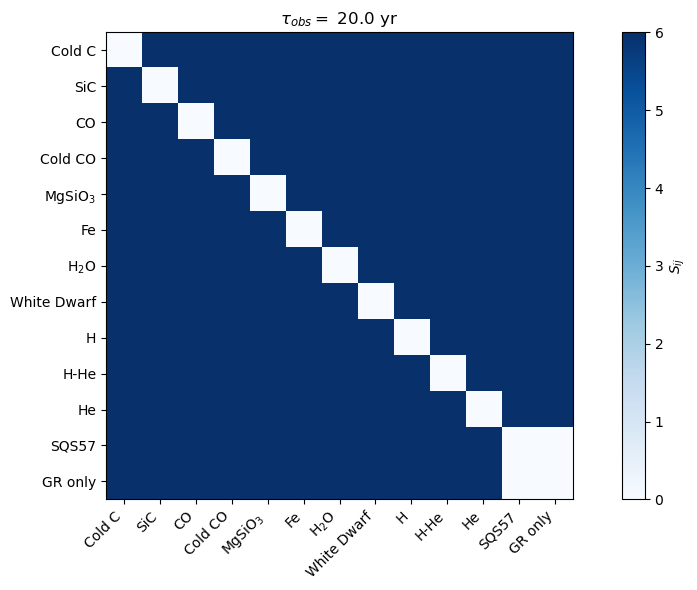}

\vspace{0.8em}

\textbf{J0636$+$5128b}

\vspace{0.3em}

\includegraphics[width=0.32\textwidth]{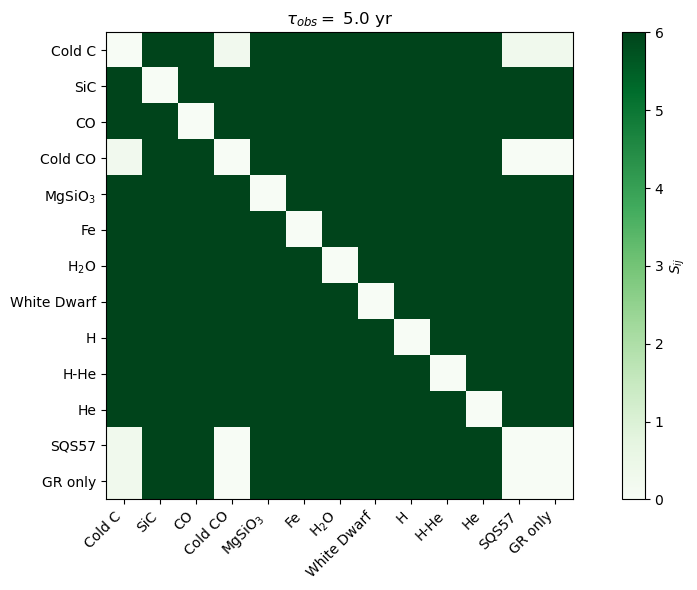}
\hfill
\includegraphics[width=0.32\textwidth]{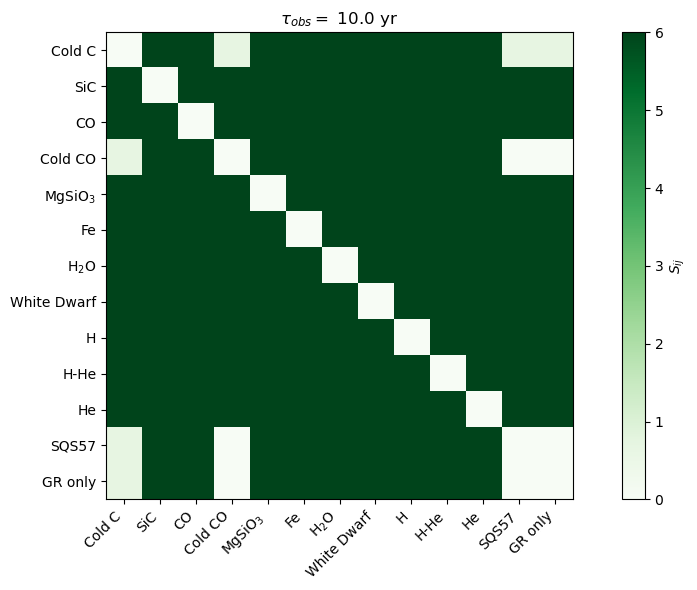}
\hfill
\includegraphics[width=0.32\textwidth]{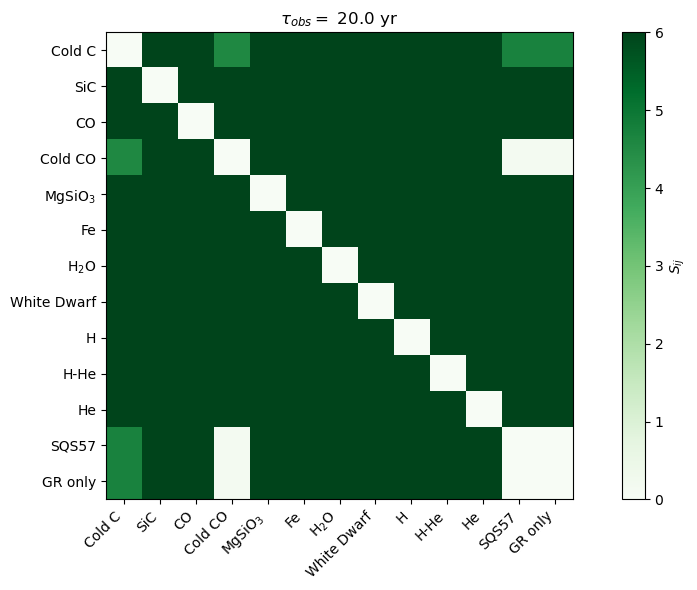}

\caption{$S_{ij}$ values for 5, 10, and 20 years of simulated observations of the J1719-1438b (top) and J0636+5128b (bottom) systems. Most EOS considered in this study have predicted $\dot\omega_{\rm tot}$ values which differ significantly enough to have the difference in their respective values greatly exceed the uncertainty on our recovered apsidal motion after only a few years of observations. Other EOS, like Cold CO, are difficult to distinguish from certain bare strange quark star models (B=57) and the GR-only precession prediction until closer to 10 or 20 years of consistent observation.}
\label{fig:sij}
\end{figure*}

%Given the lack of a significant measurement of $\dot\omega_{tot}$ in either of our target timing systems J1719-1438 and J0636+5128, we can therefore already rule out companions governed by the specific CO, SiC, MgSiO3, and Fe EOS examined here via this method. We can also rule out H-He, He, and H2O planets based on these EOS requiring large radii which would overflow their respective hosts' Roche Lobe constraints and lead to rapid tidal disruption.

For a broad class of more exotic models, small radii highly suppress Newtonian contributions, leading to an apsidal precession that is practically indistinguishable from a GR-only model. This means that it is unrealistic to expect pulsar timing to be able to distinguish between exotic companion models, at least in our methodology. If there is, in actuality, more flexibility in the possible radius of exotic candidates like strange quark stars at the mass of an ultra-light pulsar companion, it is possible such a system could obtain more substantial Newtonian precession terms. As such, a statistically robust measurement of $\dot\omega_{\rm tot}$ which deviates slightly from $\dot\omega_{\rm GR}$ could be seen as evidence that that pulsar companion is composed of some exotic state of matter.

It is worth noting that ``diamond planets'' made up of the dense carbon cores of white dwarfs are not explicitly studied here, although we take the various dense carbonaceous EOS to be proxies for their general behavior. These types of planets are often cited as a likely candidate for the identities of particularly dense pulsar companions \citep{Bailes_2011}. Our study of other metallic EOS indicate that such systems should produce apsidal motions that are unrealistically high, but we caution that even these EOS used to examine candidates made of normal matter are subject to non-insignificant uncertainties themselves. As mentioned previously, $\dot\omega_{\rm tidal}$ and $\dot\omega_{\rm spin}$ are highly sensitive to $R_c$, and even a 50$\%$ reduction in one's estimate of $R_{c}$ could reduce the projected $\dot\omega_{\rm tot}$ for these candidates below the current discerning power of pulsar timing data. As such we can make no general claim regarding the ability of pulsar timing to definitively detect apsidal motion corresponding to a ``diamond planet'' or any other candidate class, and rather emphasize the need for careful study of the EOS of candidate companions, as well as for better measurements of the orbital properties of companion systems.

%%%%%%%%%%%%%%%%%%%%%%%%%%%%%%%%%%%%%%%%%%%%%%%%%%%%%%%%%%%%%%%%%%%%%%%%%%%%%%%%%%%%
%%%%%%%%%%%%%%%%%%%%%%%%%%%%%%%%%%%%%%%%%%%%%%%%%%%%%%%%%%%%%%%%%%%%%%%%%%%%%%%%%%%%

\section{Discussion and Conclusion}\label{sec:conclusion}

We have explored a method, motivated by the feasibility of measurement, by which the internal characteristics of pulsar companions can be probed. Through modeling the apsidal motion and secular orbital decay via the use of the hydrostatic equilibrium and TOV equations, we provide a framework to falsify different EOS as compositions for our sample of low eccentricity, small orbital radii, and very high density pulsar companions. 

The EOS we study exhibit a variety of tidal characteristics, serving as a broad initial probe into the composition of these unique pulsar companions. We find that many well known terrestrial EOS can exhibit a high degree of apsidal motion with a potential level of distinguishability. Within a few decades of continued pulsar timing observations, apsidal motion can be determined to a precision of order $10^{-2}$ using an ELL1k timing model from \citet{Susobhanan_2018}, allowing for more stringent constraints on the interiors of these objects. Compact compositions were shown to return only the apsidal motion expected of a GR point mass because their small radii strongly suppress tidal effects. In the event that the apsidal motion were measured to be precisely the GR prediction with no tidal contributions, it may become necessary to rely upon future third generation gravitational wave detectors to probe companion EOS which we briefly discuss in Appendix \ref{subsec:tidal_deformability}.

With future pulsar timing measurements and possibly more JWST spectral observations on our sample like those performed by \citet{Zhang_2025} on PSR 2322-2650b, detailed EOS can be developed to fit these specific systems. The literature in which exoplanet EOS are developed often primarily focuses on the observables of mass-radius relationships \cite{Seager_2007, Podolak_2023, Wilson_2014}. Because this remains one of the most accessible means of comparing different EOS in a variety of exoplanetary systems, utilization of APSIDE provides a streamlined, straightforward framework for modeling tidal characteristics that can be used as an additional constraint to exoplanet EOS. By examining the tidal characteristics of an EOS, one opens up a more diverse landscape of comparisons that can be made, not just in pulsar companion systems.

%%%%%%%%%%%%%%%%%%%%%%%%%%%%%%%%%%%%%%%%%%%%%%%%%%%%%%%%%%%%%%%%%%%%%%%%%%%%%%%%%%%%
%%%%%%%%%%%%%%%%%%%%%%%%%%%%%%%%%%%%%%%%%%%%%%%%%%%%%%%%%%%%%%%%%%%%%%%%%%%%%%%%%%%%

\appendix
\section{Tidal Deformability and Feasibility of Measurement}\label{subsec:tidal_deformability}

\begin{figure}
\begin{center}
\mbox{\includegraphics[width = .45\textwidth]{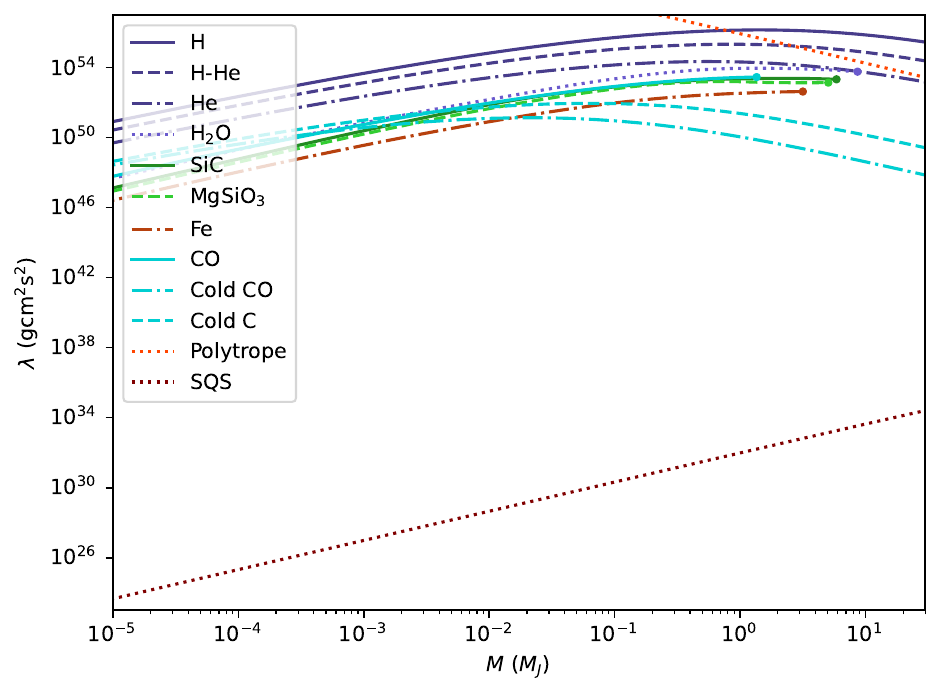}}\quad
\includegraphics[width = .45\textwidth]{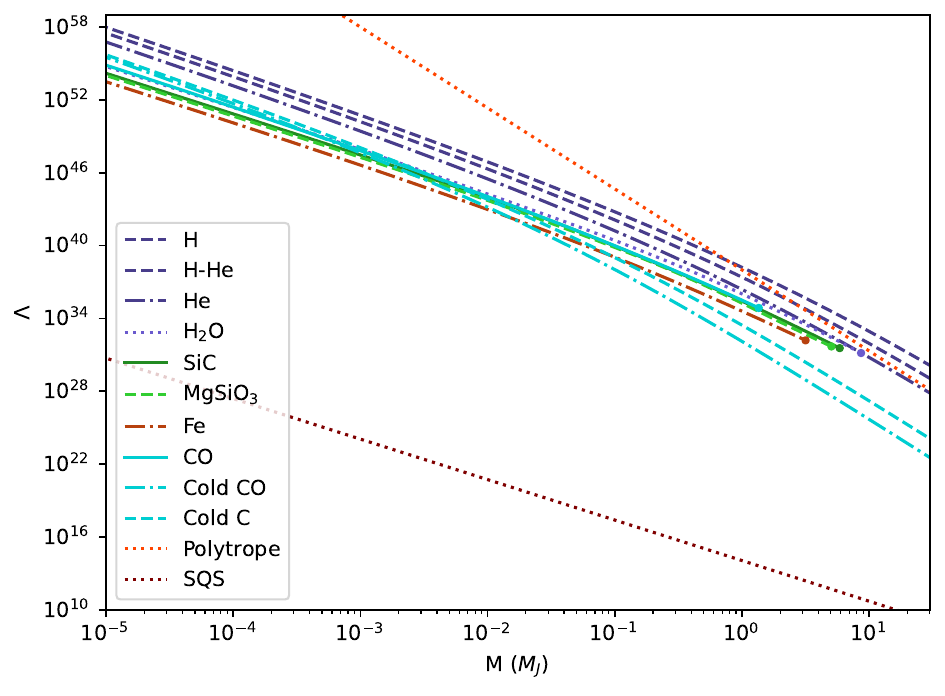}
\caption{\textit{Left:} Plot of the the relationship between mass and tidal deformabilities for our host of EOS, given by Equation \eqref{eq:tidallambda}. \textit{Right:} The same plot as left, but for the dimensionless tidal deformability, Equation \eqref{eq:dimensionless_tidallambda}.}
\label{fig:tidal-deformabilities}
\end{center}
\end{figure}

As discussed in \citep{Flanagan_2008}, the study of tidal deformability (Equation \eqref{eq:tidallambda} and Equation \eqref{eq:dimensionless_tidallambda}) is reliant upon clear gravitational wave signatures from a late-stage binary inspiral. Because the orbits of the pulsar companions we study are remarkably stable, observing a late-stage inspiral of one of these systems is likely a distant event. Despite this, modeling the tidal deformability of EOS provides a useful method of examining binary behavior and allows us to compare our results with previous works, giving another method of validating APSIDE. As shown in Figure \ref{fig:tidal-deformabilities}, the tidal deformability of companions composed of strange quark matter (in agreement with results from \citet{Wang_2021}), labeled SQS, are many orders of magnitude smaller than the rest of the tested compositions. The highly stable and compact nature of theorized strange quark objects lead to highly suppressed tidal deformabilities.

With more compact compositions, such as strange quark objects, it is theorized that an inspiral stage could exist without tidally disrupting the companion \citep{Geng_2015}, leading to a gravitational wave burst detectable by LIGO or future experiments like the Einstein Telescope \citep{Wang_2021}. Non-compact objects fail to survive in a late-stage inspiral and are tidally disrupted before they can produce a sufficiently strong GW signal. For a system like PSR J1719-1438b, this can be shown simply as follows. Following \citet{Geng_2015}, the tidal disruption radius is approximated by:

\begin{align}
    R_{td} \approx 5.1 \times 10^{10} \left( \frac{M_p}{1.4 \Msol} \right)^{1/3} \left( \frac{\rho_0}{10 g/cm^3} \right)^{-1/3}, 
\end{align}
which when using the canonical host pulsar mass of $M_p = 1.4\Msol$ and a companion density of 50 g/cm$^3$ (twice the minimum average density of this particular companion) gives a tidal disruption radius of $\approx 3\times10^{10}$ cm or about 0.002 AU. At this orbital radius the binary period will be $P_{b} =0.66$ hours. The GW strain is given by \citep{Geng_2015}:

\begin{align}
    h = 5.1\times10^{-23} \left( \frac{\mathcal{M}}{1 \Msol} \right)^{5/3} \left( \frac{P_b}{1~\text{hr}} \right)^{-2/3} \left( \frac{d}{10~\text{kpc}} \right)^{-1}.
\end{align}

\noindent We find that the strain for PSR J1719-1438b yields $h \approx 4.6 \times 10^{-25}$ at a frequency of $ 8.4\times10^{-4}$ Hz \citep{Blachier_2023}, far below the range detectable by LIGO. Hence, the recovery of a strong gravitational wave signal from one of these systems would be indicative of a more compact, exotic composition for its mass range. A measurement of a GW burst would provide an immediate means of falsifying many of the non-compact compositions we study as they are not sufficiently dense enough to produce a measurable GW strain and frequency, much like how measuring an apsidal precession exceeding the GR contribution is a quick method of falsifying exotic or compact compositions like strange quark matter objects.

%%%%%%%%%%%%%%%%%%%%%%%%%%%%%%%%%%%%%%%%%%%%%%%%%%%%%%%%%%%%%%%%%%%%%%%%%%%%%%%%%%%%
\begin{acknowledgments}
LCM thanks Ryan Lang and Tanja Hinderer for advice in the early stages of developing APSIDE and Barinov M.V. for a correction to a system name. This work is partly supported by the U.S.\ Department of Energy grant number de-sc0010107 (SP).
\end{acknowledgments}

%%%%%%%%%%%%%%%%%%%%%%%%%%%%%%%%%%%%%%%%%%%%%%%%%%%%%%%%%%%%%%%%%%%%%%%%%%%%%%%%%%%%

\bibliography{bibliography.bib}
\bibliographystyle{aasjournalv7}

\end{document}